\documentclass[twocolumn,superscriptaddress,
showpacs,preprintnumbers,amsmath,amssymb]{revtex4}

\usepackage{graphicx}
\usepackage{amssymb,amsmath,amsthm}

\theoremstyle{definition}

\newcommand{\bra}[1]{{\left\langle #1 \right|}}
\newcommand{\ket}[1]{{\left| #1 \right\rangle}}

\newcommand{\T}{\mbox{$\mathrm{tr}$}}

\begin{document}

\title{Concurrence of assistance and Mermin inequality on three-qubit pure states}

\author{Dong Pyo Chi}
\affiliation{
 Department of Mathematical Sciences,
 Seoul National University, Seoul 151-742, Korea
}
\author{Kabgyun Jeong}
\affiliation{
 Nano Systems Institute (NSI-NCRC),
 Seoul National University, Seoul 151-742, Korea
 }
\author{Taewan Kim}
\affiliation{
 Department of Mathematical Sciences,
 Seoul National University, Seoul 151-742, Korea
}
\author{Kyungjin Lee}
\affiliation{
 Department of Mathematical Sciences,
 Seoul National University, Seoul 151-742, Korea
}
\author{Soojoon Lee}
\affiliation{
 Department of Mathematics and Research Institute for Basic Sciences,
 Kyung Hee University, Seoul 130-701, Korea
}
\date{\today}

\begin{abstract}
We study a relation between the concurrence of assistance and
the Mermin inequality on three-qubit pure states.
We find that
if a given three-qubit pure state has
the minimal concurrence of assistance greater than $1/2$
then the state violates some Mermin inequality.
\end{abstract}

\maketitle

Bell-inequality violation in quantum mechanics tells us that
quantum correlations are quite different from classical correlations.
In the case of two-qubit states,
the Clauser-Horne-Shimony-Holt (CHSH) inequality~\cite{CHSH}
is a well-known Bell inequality,
and has an important property that
any two-qubit pure state violates the CHSH inequality
if and only if it is entangled.
In particular,
there exists an explicit relation
between the degree of the CHSH-inequality violation and the amount of entanglement
for two-qubit pure states~\cite{Gisin91}.
This shows that entanglement of pure states in the two-qubit system
can be certainly detected by employing the Bell inequality,
and the Bell-inequality violation can be exactly determined 
according to the amount of entanglement
for two-qubit pure states.
On this account,
there have been a lot of research works to attempt
to generalize the explicit relation into the multiqubit pure states~\cite{SG01,ZBLW,EB,CWKO,GSDRS}.

We here consider the Mermin inequality~\cite{Mermin} for three-qubit pure states,
which is a natural generalization of the CHSH inequality:
Let $\mathcal{B}_M$ be the operator defined as
\begin{eqnarray}
\mathcal{B}_M&=&
\vec{a}_1\cdot\vec{\sigma}\otimes\vec{a}_2\cdot\vec{\sigma}\otimes\vec{a}_3\cdot\vec{\sigma}
-\vec{a}_1\cdot\vec{\sigma}\otimes\vec{b}_2\cdot\vec{\sigma}\otimes\vec{b}_3\cdot\vec{\sigma}
\nonumber\\
&&-\vec{b}_1\cdot\vec{\sigma}\otimes\vec{a}_2\cdot\vec{\sigma}\otimes\vec{b}_3\cdot\vec{\sigma}
-\vec{b}_1\cdot\vec{\sigma}\otimes\vec{b}_2\cdot\vec{\sigma}\otimes\vec{a}_3\cdot\vec{\sigma},
\nonumber\\
\label{eq:M_operator}
\end{eqnarray}
where $\vec{a}_j$ and $\vec{b}_j$ are unit vectors in $\mathbb{R}^3$,
and $\vec{\sigma}=(\sigma_1,\sigma_2,\sigma_3)$ is the vector of the Pauli matrices.
Then for a given three-qubit pure state $\ket{\psi}$, the Mermin inequality is
\begin{equation}
\left|\bra{\psi} \mathcal{B}_M\ket{\psi}\right|\le 2.
\label{eq:Mermin_ineq}
\end{equation}

For the generalized Greenberger-Horne-Zeilinger (GHZ) states,
\begin{equation}
\ket{\psi_{\mathrm{GHZ}}}=\cos{\phi}\ket{000}+\sin{\phi}\ket{111},
\label{eq:GGHZ}
\end{equation}
it was numerically shown in Ref.~\cite{SG01} that
the state $\ket{\psi_{\mathrm{GHZ}}}$ violates a Mermin inequality
if and only if $\sin{2\phi}>1/2$.
This result implies that there exists a relation between the Mermin-inequality violation
and the amount of entanglement for three-qubit pure states,
since $\sin{2\phi}$ may represent the degree of entanglement in the state $\ket{\psi_{\mathrm{GHZ}}}$.
Then one could naturally ask whether the same result can be obtained for any three-qubit pure state.

In order to answer this question,
the proper quantity such as the value $\sin{2\phi}$ in the generalized GHZ states
should be defined for general three-qubit pure states, and
it should be investigated whether the Mermin inequality is violated,
whenever the quantity is greater than some constant.
In this paper,
we consider the {\it concurrence of assistance} (CoA)~\cite{LVV} as such a quantity,
and examine a relation between the CoA and the Mermin-inequality violation
for several classes of three-qubit pure states including
the generalized GHZ states, the states in the W class,
and some coherent superpositions of well-known three-qubit pure states.

As a consequence, we analytically show that
if a three-qubit pure state in those classes has
the minimal CoA greater than $1/2$
then the state violates a Mermin inequality,
and furthermore find that
our result can be generalized into all three-qubit pure states
by exploiting the numerical work in Ref.~\cite{EB}.

We first take account of two simple but important measures of entanglement,
the concurrence~\cite{Wootters} and the CoA.
The concurrence, $\mathcal{C}$, is defined as follows:
For a pure state $\ket{\phi}_{12}$ in $2\otimes d$ quantum systems ($d\ge 2$),
it is defined as
$\mathcal{C}(\ket{\phi}_{12}\bra{\phi})=\sqrt{2(1-\T\rho_1^2)}=2\sqrt{\det\rho_1}$,
where $\rho_1=\T_2\ket{\phi}_{12}\bra{\phi}$.
For any mixed state $\rho_{12}$, it is defined as
\begin{equation}
\mathcal{C}(\rho_{12})=\min \sum_{k} p_k
\mathcal{C}(\ket{\phi_k}_{12}\bra{\phi_k}),
\label{eq:concurrence}
\end{equation}
where the minimum is taken over its all possible decompositions,
$\rho_{12}=\sum_k p_k \ket{\phi_k}_{12}\bra{\phi_k}$.
The CoA, $\mathcal{C}^a$, is also defined in the similar way:
For a pure state $\ket{\phi}_{12}$,
$\mathcal{C}^a(\ket{\phi}_{12}\bra{\phi})\equiv\mathcal{C}(\ket{\phi}_{12}\bra{\phi})$.
For a mixed state $\rho_{12}$, it is defined as
\begin{equation}
\mathcal{C}^a(\rho_{12})=\max \sum_{k} p_k
\mathcal{C}(\ket{\phi_k}_{12}\bra{\phi_k}),
\label{eq:concurrence}
\end{equation}
where the maximum is taken over all possible decompositions of $\rho_{12}$.

We remark that
the CoA is an entanglement monotone on three-qubit pure states~\cite{GMS}.
Thus, even though the definitions of the two entanglement measures are quite similar,
the CoA can be thought of as a measure of entanglement on tripartite pure states,
while the concurrence is a good measure of bipartite entanglement.
Our aim in this paper is to define an appropriate measure of entanglement for three-qubit pure states,
and to investigate how the entanglement measure is related to the Mermin-inequality violation.
Hence, the CoA may be one of good candidates for such a tripartite entanglement measure.

Furthermore, it was known in Ref.~\cite{CCJKKL}
that, for any three-qubit pure state $\ket{\psi}_{123}$,
there exists the so-called monogamy equality
in terms of the concurrence and the CoA as follows:
\begin{equation}
\mathcal{C}_{1(23)}^2=\mathcal{C}_{12}^2+\left(\mathcal{C}_{13}^a\right)^2,
\label{eq:ME222}
\end{equation}
where $\mathcal{C}_{1(23)}=\mathcal{C}(\ket{\psi}_{1(23)}\bra{\psi})$,
$\mathcal{C}_{12}=\mathcal{C}(\T_3\ket{\psi}_{123}\bra{\psi})$, and
$\mathcal{C}_{13}^a=\mathcal{C}^a(\T_2\ket{\psi}_{123}\bra{\psi})$.
Thus, for distinct $i$ and $j$ in $\{1,2,3\}$, we clearly have the equalities
$\mathcal{C}_{ij}^a=\sqrt{\tau+\mathcal{C}^2_{ij}}$,
where $\tau$ is called the three-tangle~\cite{CKW,DVC},
defined as
\begin{equation}
\tau=\mathcal{C}^2_{1(23)}-\mathcal{C}^2_{12}-\mathcal{C}^2_{13},
\label{eq:tangle}
\end{equation}
and is known as an entanglement measure to distinguish
the GHZ class from the W class~\cite{DVC}.
Here, the GHZ class and the W class are
the sets of all pure states with genuine three-qubit entanglement
equivalent to the GHZ state~\cite{GHZ},
\begin{equation}
\ket{\mathrm{GHZ}}=\frac{1}{\sqrt{2}}\left(\ket{000}+\ket{111}\right),
\label{eq:GHZ}
\end{equation}
under stochastic local operations and classical communication (SLOCC),
and equivalent to the W state,
\begin{equation}
\ket{\mathrm{W}}=\frac{1}{\sqrt{3}}\left(\ket{001}+\ket{010}+\ket{100}\right),
\label{eq:W}
\end{equation}
under SLOCC, respectively.

Now, we consider the Schmidt decomposition of three-qubit pure states as follows~\cite{AACJLT}:
\begin{eqnarray}
\ket{\psi}_{123}&=&\lambda_0\ket{000}_{123}
+\lambda_1 e^{\iota\theta}\ket{100}_{123}
+\lambda_{2}\ket{101}_{123}\nonumber\\
&&+\lambda_{3}\ket{110}_{123}+\lambda_{4}\ket{111}_{123},
\label{eq:Schmidt3}
\end{eqnarray}
where $\iota=\sqrt{-1}$, $0\le\theta\le\pi$, $\lambda_j\ge 0$, and $\sum_j\lambda_j^2=1$.
Thus, in order to calculate the CoAs for three-qubit pure states,
it suffices to consider the states in Eq.~(\ref{eq:Schmidt3}).
By somewhat tedious but straightforward calculations,
we obtain the following results on the CoAs $\mathcal{C}_{ij}^a$ for $\ket{\psi}_{123}$:
\begin{eqnarray}
\mathcal{C}_{12}^a&=&2\lambda_0\sqrt{\lambda^2_3+\lambda^2_4},\nonumber\\
\mathcal{C}_{23}^a&=&2\sqrt{\lambda^2_0\lambda^2_4+\lambda^2_1\lambda^2_4+\lambda^2_2\lambda^2_3
-2\lambda_1\lambda_2\lambda_3\lambda_4\cos\theta},\nonumber\\
\mathcal{C}_{31}^a&=&2\lambda_0\sqrt{\lambda^2_2+\lambda^2_4}.
\label{eq:calculation_CoA}
\end{eqnarray}
Let $\mathcal{C}^a_{\mathrm{min}}=\min\{\mathcal{C}_{12}^a,\mathcal{C}_{23}^a,\mathcal{C}_{31}^a\}$.
Then $\mathcal{C}^a_{\mathrm{min}}$ is called the minimal CoA,
which is the very entanglement measure relevant to our purpose.
Our claim is that, for a given three-qubit pure state,
if its minimal CoA is greater than 1/2 then
there exists a Mermin inequality which the state violates.

For the generalized GHZ states $\ket{\psi_\mathrm{GHZ}}$ in Eq.~(\ref{eq:GGHZ}),
it is easy to calculate
that $\mathcal{C}^a_{\mathrm{min}}=\sin2\phi$
by Eqs.~(\ref{eq:calculation_CoA}).
Take $\vec{a}_j=(1,0,0)$ and $\vec{b}_j=(0,1,0)$ for all $j=1, 2, 3$.
Then the Mermin inequality in Eq.~(\ref{eq:Mermin_ineq})
becomes $4\sin2\phi\le 2$.
Hence, we can clearly obtain that
if the generalized GHZ state has the minimal CoA greater than 1/2
then the state violates a Mermin inequality.
In other words, our claim is true for the generalized GHZ states.

We now take the W class into account.
It is known in~\cite{DVC,ABLS}
that any state $\ket{\psi_\mathrm{W}}$ in the W class can be written as
\begin{equation}
\ket{\psi_\mathrm{W}}=\lambda_0\ket{000}+\lambda_{1}\ket{100}
+\lambda_{2}\ket{101}+\lambda_{3}\ket{110},
\label{eq:Wclass}
\end{equation}
which has the simpler Schmidt decomposition than the general one in Eq.~(\ref{eq:Schmidt3}),
since its three-tangle is zero.
In this case,
we take $\vec{a}_j=(0,0,1)$ for all $j=1, 2, 3$,
$\vec{b}_1=(-1,0,0)$, and $\vec{b}_2=\vec{b}_3=(1,0,0)$.
Then the Mermin inequality in Eq.~(\ref{eq:Mermin_ineq})
becomes
\begin{equation}
\lambda_0^2-\lambda_1^2+\lambda_2^2+\lambda_3^2
+2\lambda_0\lambda_2+2\lambda_2\lambda_3+2\lambda_3\lambda_0\le 2.
\label{eq:W_Mermin}
\end{equation}
By Eqs.~(\ref{eq:calculation_CoA}), the minimal CoA, $\mathcal{C}^a_{\mathrm{min}}$, for the W class
can be readily calculated as
\begin{equation}
\mathcal{C}^a_{\mathrm{min}}
=2\min\{\lambda_0\lambda_2,\lambda_2\lambda_3,\lambda_3\lambda_0\}.
\label{eq:CoA_W}
\end{equation}
Since $\lambda_i^2+\lambda_j^2\ge 2\lambda_i\lambda_j$
and $\lambda_0^2+\lambda_1^2+\lambda_2^2+\lambda_3^2=1$,
we can obtain the following inequalities:
\begin{eqnarray}
2(1-\lambda_1^2)&=&
2(\lambda_0^2+\lambda_2^2+\lambda_3^2)\nonumber\\
&\ge& 2\lambda_0\lambda_2+2\lambda_2\lambda_3+2\lambda_3\lambda_0\nonumber\\
&\ge& 3 \mathcal{C}^a_{\mathrm{min}}.
\label{W_ineq}
\end{eqnarray}
Thus, if $\mathcal{C}^a_{\mathrm{min}}>1/2$ then
$\lambda_1^2<1/4$, and hence
the left-hand side of the inequality in~(\ref{eq:W_Mermin})
is greater than 2, that is,
its Mermin inequality is violated.
Therefore, our claim is also true for the W class.

We now consider three coherent superpositions of well-known states.
First, let us see a coherent superposition of
the generalized GHZ state and a separable state $\ket{101}$,
\begin{equation}
\ket{{\psi_\mathrm{GHZ}:\mathrm{S}}}=\sqrt{1-p}\ket{\psi_\mathrm{GHZ}}+\sqrt{p}\ket{101},
\label{eq:GHZ_S}
\end{equation}
where $0< p< 1$.
Taking account of the Mermin inequality used in the case of the generalized GHZ states,
we can show that
the Mermin inequality is violated if and only if $4(1-p)\sin2\phi>2$,
and that the minimal CoA for $\ket{{\psi_\mathrm{GHZ}:\mathrm{S}}}$ equals $(1-p)\sin2\phi$ 
by Eqs.~(\ref{eq:calculation_CoA}).
Hence, it is clear that
if the minimal CoA for the state $\ket{{\psi_\mathrm{GHZ}:\mathrm{S}}}$ is more than 1/2
then it violates the same Mermin inequality as the inequality for the generalized GHZ states,
and vice versa.

The second coherent superposition which we deal with
is a superposition of the W state and a separable state $\ket{000}$,
\begin{equation}
\ket{\mathrm{W:S}}=\sqrt{1-p}\ket{\mathrm{W}}+\sqrt{p}\ket{000},
\label{eq:W_S}
\end{equation}
where $0< p< 1$.
We now take $\vec{a}_j=(0,0,-1)$ and $\vec{b}_j=(1,0,0)$ for all $j=1, 2, 3$.
Then the Mermin inequality becomes $3-4p\le 2$, that is, $p\ge 1/4$.
It can be obtained from simple calculations~\cite{simple}
that its minimal CoA is $2(1-p)/3$.
Thus, the minimal CoA for the state $\ket{\mathrm{W:S}}$ is greater than 1/2
if and only if the Mermin inequality
with respect to $\vec{a}_j=(0,0,-1)$ and $\vec{b}_j=(1,0,0)$
is violated.

\begin{figure}
\includegraphics[angle=-90,width=.95\linewidth]{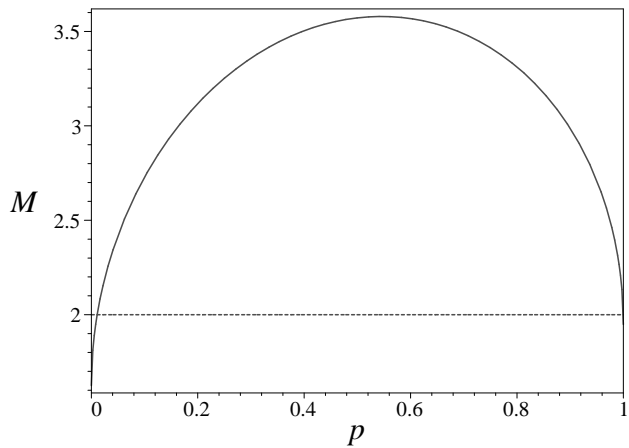}
\caption{\label{Fig:GHZ_W}
The left-hand side $M(p)$ of the inequality in~(\ref{eq:GHZ_W_Mermin1})
is more than two for $0.011\le p\le 0.999$.}
\end{figure}

Let us now deal with the coherent superposition of the GHZ state and the W state,
\begin{equation}
\ket{\mathrm{GHZ:W}}=\sqrt{1-p}\ket{\mathrm{GHZ}}+\sqrt{p}\ket{\mathrm{W}},
\label{eq:W_S}
\end{equation}
where $0< p< 1$.
Then it follows from direct computations that
for all $0< p< 1$ the states $\ket{\mathrm{GHZ:W}}$ have the minimal CoA more than 1/2.
Thus, it suffices to show that the state violates some Mermin inequality for each $0<p<1$.
We here use three Mermin inequalities to show the violation,
according to the value of the parameter $p$.
We first consider the Mermin inequality with respect to \
$\vec{a}_j=(1/2,0,-\sqrt{3}/2)$ and $\vec{b}_j=(0,1,0)$ for all $j=1, 2, 3$.
Then the Mermin inequality for the states $\ket{\mathrm{GHZ:W}}$ becomes
\begin{eqnarray}
\frac{1}{8}\left(3\sqrt{2}(5+\sqrt{3})\sqrt{p(1-p)}+13(1-p)+9\sqrt{3}p\right)
\le 2.\nonumber \\
\label{eq:GHZ_W_Mermin1}
\end{eqnarray}
Let $M(p)$ be the left-hand side of the inequality in~(\ref{eq:GHZ_W_Mermin1}).
Then it can be shown that
$M(p)$ is greater than two if $0.011\le p\le 0.999$,
and hence the Mermin inequality is violated in this case,
which is depicted in FIG.~\ref{Fig:GHZ_W}.
Furthermore, it can be readily shown that
the states $\ket{\mathrm{GHZ:W}}$ for $0<p<1/2$ and $1>p>(9+\sqrt{21})/15 \simeq 0.9055$
violate the Mermin inequalities taken in the cases of the states $\ket{\psi_\mathrm{GHZ}}$
and the states $\ket{\mathrm{W:S}}$, respectively.
Therefore, our claim also holds for the states $\ket{\mathrm{GHZ:W}}$.

In addition to our analytical results,
our claim can be generalized into any three-qubit pure states,
by exploiting the numerical work of Emary and Beenakker in Ref.~\cite{EB}.
In their work, an entanglement measure $\sigma$ was defined as
\begin{equation}
\sigma\equiv \min\left(\frac{\mathcal{C}_{X(YZ)}^2+\mathcal{C}_{Y(XZ)}^2}{2}-\mathcal{C}_{XY}^2\right),
\label{eq:sigma}
\end{equation}
where the minimization is over the permutations $X, Y, Z$ in $\{1, 2, 3\}$.
Then we can obtain the simple relation among
the three-tangle, the minimal CoA, and this measure of entanglement $\sigma$ as follows:
\begin{equation}
0\le \tau\le \left(\mathcal{C}^a_{\mathrm{min}}\right)^2\le \sigma \le 1.
\label{eq:relation}
\end{equation}
Since their numerical result shows the following inequality
\begin{equation}
\sigma\le \frac{\left|\bra{\psi} \mathcal{B}_M\ket{\psi}\right|^2}{16}
\label{eq:numerical_ineq}
\end{equation}
for numbers of three-qubit pure states $\ket{\psi}$,
this implies that our claim is numerically true for general three-qubit pure states.

In summary, we have studied 
a relation between the CoA and the Mermin-inequality violation 
for several classes of three-qubit pure states,
and have obtained an analytical result 
that if a three-qubit pure state in those classes has
the minimal CoA greater than $1/2$
then the state violates some Mermin inequality.
Furthermore, we have also found that 
our result numerically holds for any three-qubit pure states.

This work was supported by the IT R\&D program of MKE/IITA
(2008-F-035-02,
Development of key technologies for commercial quantum cryptography communication system).
D.P.C. was supported by the Korea Science and Engineering Foundation
(KOSEF) grant funded by the Korea government (MOST) (No.~R01-2006-000-10698-0),
and S.L. was supported by Basic Science Research Program
through the National Research Foundation of Korea (NRF)
funded by the Ministry of Education, Science and Technology (Grant No.~2009-0076578).



\begin{thebibliography}{1}
\bibitem{simple}
If $\sigma_x\otimes I\otimes I$ is applied to the state $\ket{\mathrm{W:S}}$,
the resulting state is of the form in Eq.~(\ref{eq:Schmidt3}),
and hence the minimal CoA for the state can be calculated by Eqs.~(\ref{eq:calculation_CoA}).
\end{thebibliography}
\end{document}